\documentclass[11pt,twoside]{article}

\usepackage{asp2004}
\usepackage{epsf}
\usepackage{psfig}
\usepackage{lscape}
\usepackage{natbib}

\begin{document}
\title{Evolution of Gamma-Ray Burst Progenitors at Low Metallicity}   %%% Fill in title
\author{S.-C. Yoon$^{1}$ and N. Langer$^{2}$}   %%% Fill in author names
\affil{1. Astronomical Institute "Anton Pannekoek", University of Amsterdam,  Kruislaan 403, 1098 SJ, Amsterdam, The Netherlands \\
       2. Astronomical Institute, Utrecht University, Princetonplein 5, 3584 CC, Utrecht, The Netherlands
}    %%% Fill in author affiliations

\begin{abstract} %%% Abstract to run on from here.
Despite the growing evidence that long Gamma-Ray
Bursts (GRBs) are associated with deaths of Wolf-Rayet stars,
the evolutionary path of massive stars to GRBs and the exact nature
of GRB progenitors remained poorly known.
However, recent massive star evolutionary models indicate
that --- for sufficiently low metallicity --- 
initially very rapidly rotating stars can
satisfy the conditions for collapsar formation.
Even though magnetic torques are included in these models,
a strong core spin-down is avoided
through quasi-chemically homogeneous evolution induced by rotational 
mixing.  Here, we explore for which initial mass and spin-range
single stars of $Z=Z_{\odot}/20$ are expected to produce GRBs.
We further find a dichotomy in the chemical structure of GRB progenitors,
where lower initial masses end their lives with a massive
helium envelope which still contains some amounts of hydrogen,
while higher initial masses explode with C/O-dominated 
hydrogen-free atmospheres. 
\end{abstract}

%%% MAIN BODY OF TEXT GOES HERE. CONSULT "INSTRUCTIONS FOR AUTHORS USING
%%% LATEX2E MARKUP", SECTIONS 2.3-2.6 FOR HELP WITH EQUATIONS, FIGURES,
%%% AND TABLES.

\section{Introduction}   %%% Top level section head (remove "%" symbol)

Long gamma-ray bursts (GRBs) are believed to originate from 
rapidly rotating massive Wolf-Rayet stars
(see Woosley \& Heger~\citeyear{Woosley04} for a review).
Interestingly, recent observations indicate that GRBs occur 
preferentially in metal poor environments (Fynbo et al.~\citeyear{Fynbo03}; Conselice et al.~\citeyear{Conselice05}; 
Gorosabel et al.~\citeyear{Gorosabel05}; Chen et al.~\citeyear{Chen05}; Starling et al.~\citeyear{Starling05a}).
However, it has been questioned whether metal poor single stars are able to produce 
rapidly rotating WR stars as GRB progenitors,
based on two reasons. First, stellar evolution models which include magnetic torques
indicate that the core loses too much angular momentum during the giant phase
to produce collapsar and GRBs (Heger, Woosley \& Spruit~\citeyear{Heger05}; Petrovic et al.~\citeyear{Petrovic05}). 
Second, the lower the metallicity, the more difficult is
the removal of the hydrogen envelope -- without which 
jets from the central engine could not escape from the star -- 
even from very massive stars (see, however, Meynet et al.~\citeyear{Meynet05b}).

Two recent independent studies by Woosley \& Heger (\citeyear{Woosley05}) and
Yoon \& Langer (\citeyear{Yoon05})
give a plausible solution to this problem.
They considered so-called homogeneous evolution
(Maeder~\citeyear{Maeder87}),
where rotationally induced chemical mixing induces
quasi-homogeneity of the chemical composition of the star throughout core hydrogen burning.
In this case, single stars can become Wolf-Rayet stars
without the need of stellar wind mass loss, 
and they avoid the giant phase that otherwise would
cause a significant decrease of the core angular momentum 
by magnetic torques.
The above mentioned authors showed that
such evolution can actually lead to the retention of enough angular momentum
in the core to produce GRBs,
if metallicity is sufficiently low ($Z \la Z_\odot/10$). 
Here we  present stellar evolution models at $Z=0.001$
which include rotation and magnetic toques (Spruit~\citeyear{Spruit02}), and systematically investigate
in which conditions stellar evolution can lead to GRBs, via such an evolutionary path.

\section{Methods}

The stellar models are calculated with a hydrodynamic
stellar evolution code, which includes the effect of
rotation on the stellar structure, rotationally
induced chemical mixing, and the transport
of angular momentum by magnetic torques 
(see Petrovic et al.~\citeyear{Petrovic05} and references therein). 
We follow  Kudritzki et al. (\citeyear{Kudritzki89})  
for stellar wind mass loss of hydrogen rich stars.
Wolf-Rayet wind mass loss rates are adopted following Hamann et al. (\citeyear{Hamann95}), but 
reduced by a factor of 10, which corresponds to the recent estimates by Vink \& de Koter (\citeyear{Vink05}). 
The effect of the enrichment of CNO elements at the stellar surface on the WR wind mass loss rate
is also considered such that, with a given surface condition, 
WC stars with $X_\mathrm{CNO} = 0.5$ and $X_\mathrm{He}=0.5$
have about 10 times higher mass loss rates than WN stars.
A metallicity dependence of $\dot{M} \propto Z^{0.69}$
and $\dot{M} \propto Z^{0.86}$ is adopted for hydrogen rich stars and WR stars, respectively, 
following Vink et al.~(\citeyear{Vink01}) and Vink \& de Koter~(\citeyear{Vink05}). 
We do not consider overshooting in the convective region, but employ
rather fast semi-convection with an efficiency parameter $\alpha_\mathrm{SEM} = 1.0$ 
(see Langer~\citeyear{Langer91}). Uncertainties due to these assumptions, and
their effects on the results are discussed in Yoon \& Langer (\citeyear{Yoon06}). 

\section{Evolution of massive stars at $Z=0.001$}
\begin{figure}[!t]
\begin{center}
\resizebox{0.62\hsize}{!}{\includegraphics{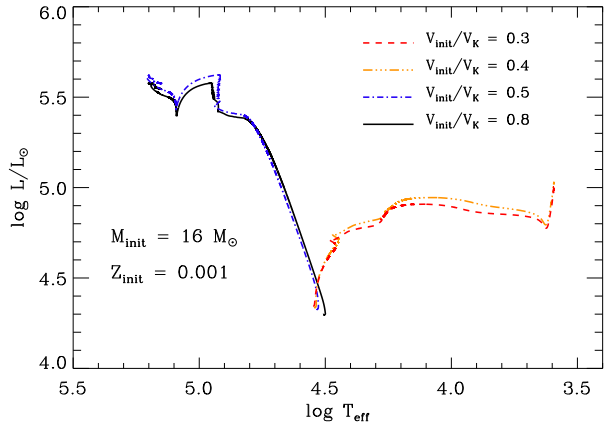}}
%\plotone{hr.eps}
\caption{Evolution of $16~\mathrm{M_\odot}$ stellar models at $Z_\mathrm{init}=0.001$, 
for different initial rotational velocities ($V_\mathrm{init}/V_\mathrm{Kepler} = 0.3, 0.4, 0.5, 0.8$), 
in the HR diagram.}\label{fig:hr}
\end{center}
\end{figure}

\begin{figure}[!t]
\resizebox{0.5\hsize}{!}{\includegraphics{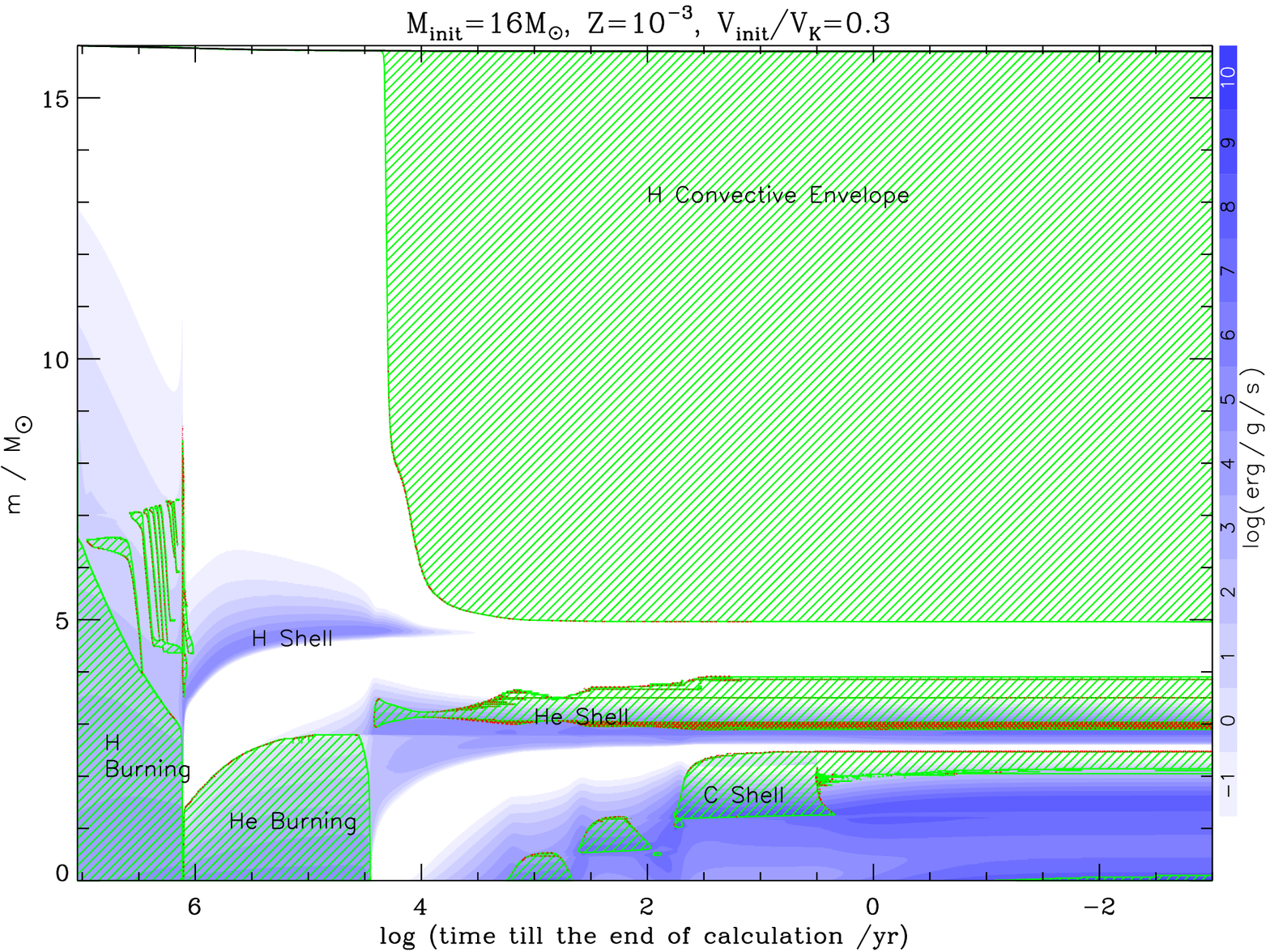}}
\resizebox{0.5\hsize}{!}{\includegraphics{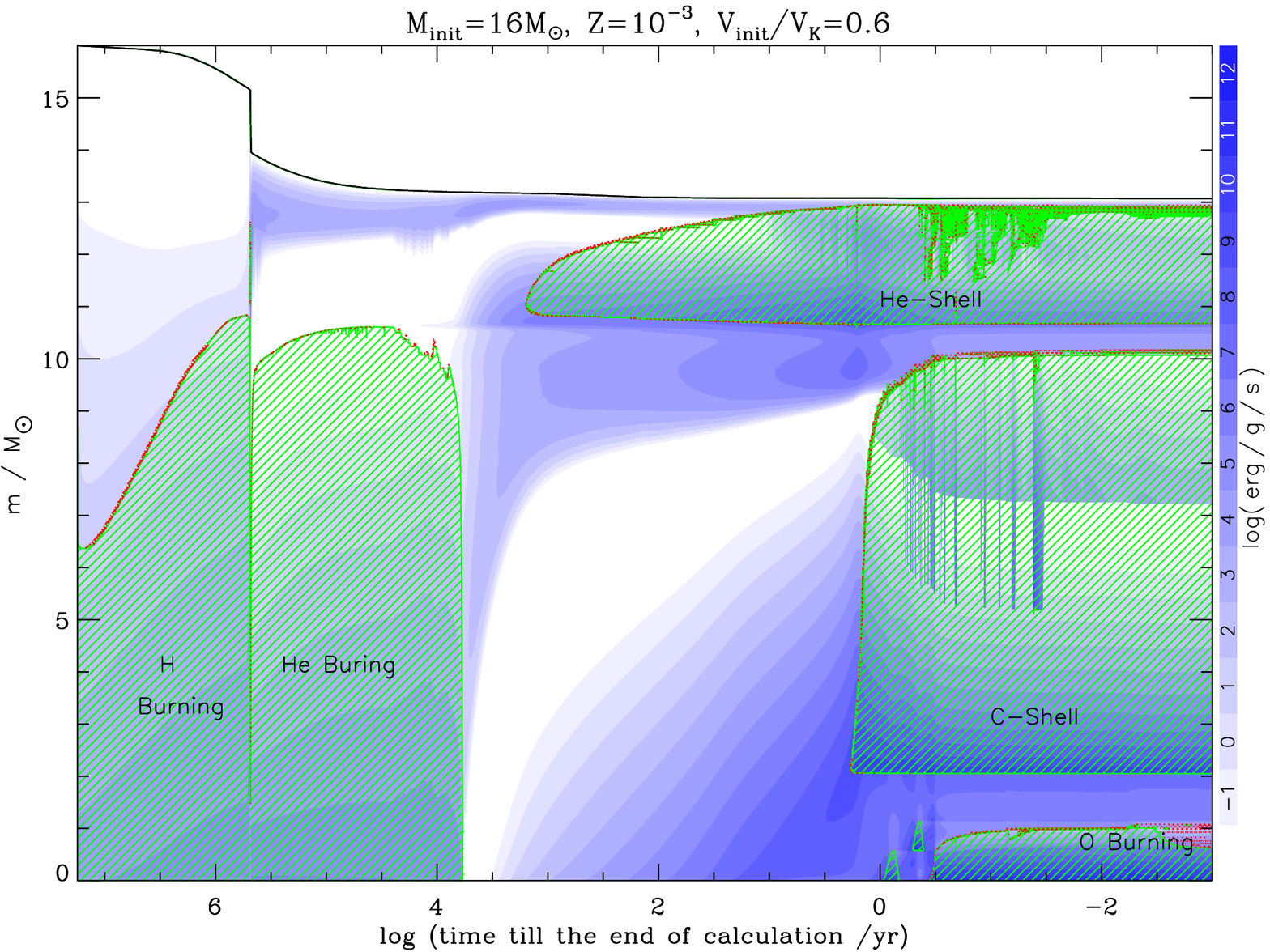}}
%\plottwo{kipp1.eps}{kipp2.eps}
\caption{Evolution of the internal structure of the model sequences of $M_\mathrm{init}=16\mathrm{M_\odot}$ and $Z=0.001$, 
with $V_\mathrm{init}/V_\mathrm{K}=0.3$ (left panel) and $V_\mathrm{init}/V_\mathrm{K}=0.5$ (right panel). 
Convective regions are hatched. Gray shading denotes nuclear energy generation. 
}\label{fig:kipp}
\end{figure}

As discussed by Maeder (\citeyear{Maeder87}), the evolution 
of rotating stars can bifurcate according to the initial spin rate.
While non-rotating, or slowly rotating stars evolve redwards, 
rapidly rotating stars evolve bluewards due to 
very efficient rotationally induced chemical mixing. 
Figure~\ref{fig:hr} illustrates this bifurcation in the H-R diagram, with
$16~\mathrm{M_\odot}$ models.
Initially slowly rotating stars ($V_\mathrm{init}/V_\mathrm{K} = 0.3~\&~0.4$)
transform into red giants with a massive extended hydrogen envelope (Fig.~\ref{fig:kipp}). 
The resulting CO core mass is about $2.8~\mathrm{M_\odot}$, and the star is expected
to explode as a Type II supernova leaving a neutron star as a remnant.
On the other hand, the initially rapidly rotating stars ($V_\mathrm{init}/V_\mathrm{K} = 0.5~\&~0.8$),
which become WR stars on the main sequence due to rotationally induced mixing, 
have only a tiny compact hydrogen envelope during the core He burning phase 
(Fig.~\ref{fig:kipp}).
Importantly, such a tiny envelope 
cannot spin down the rapidly rotating core, contrary to the
case where stars evolve into red giants (Heger et al.~\citeyear{Heger05}; Petrovic et al.~\citeyear{Petrovic05}).
Rather efficient angular momentum transport from the CO core occurs after the core He exhaustion, 
but the central core can retain enough angular momentum to produce collapsar 
(Woosley \& Heger~\citeyear{Woosley05}; Yoon \& Langer~\citeyear{Yoon05}).
The CO core mass at the final stage is about $10.4~\mathrm{M_\odot}$, which is large
enough to form a black hole.

\begin{figure}[!t]
\begin{center}
\resizebox{0.9\hsize}{!}{\includegraphics{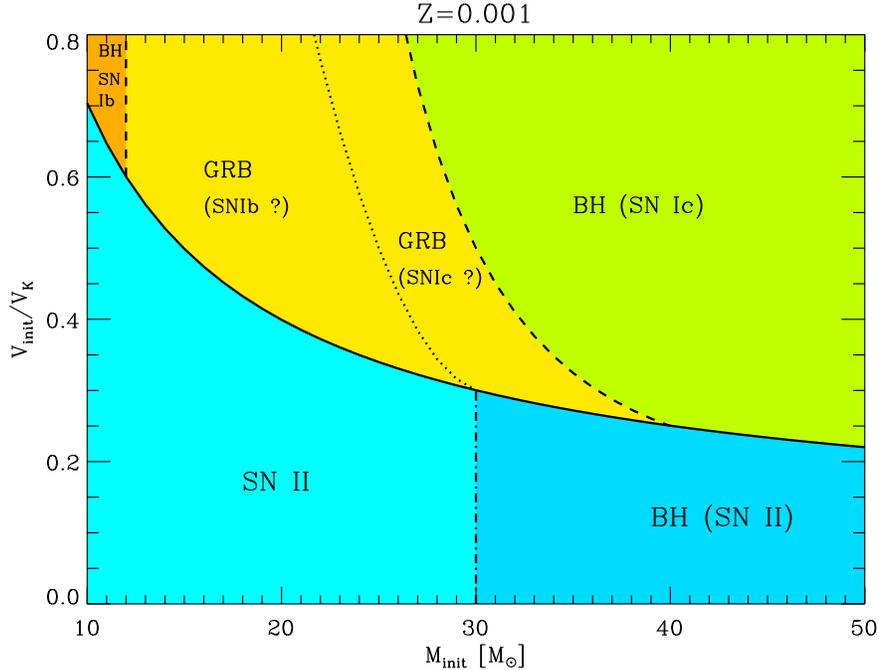}}
\caption{Final fate of rotating massive stars at $Z=0.001$, in the plane of initial mass and 
initial fraction of the Keplerian value of the equatorial rotational velocity.
The solid line divides the plane into two parts, where stars evolve quasi-chemically homogeneous
above the line, while they evolve into the classical core-envelope structure below the line.
Between the dashed lines is the region of quasi-homogeneous evolution 
where the core mass, core spin and stellar radius are compatible with the collapsar 
model for GRB production, while to both sides of it black holes are formed but
the core spin is insufficient. 
This GRB production region is divided into two parts, where GRB progenitors
do or do not possess a thick helium envelope
(i.e. $\Delta M_\mathrm{He} \ga 2.0 \mathrm{M_\odot}$).
The dashed-dotted line in the region of non-homogeneous evolution 
separates Type II supernovae (SN II; left) and black hole (BH; right) formation,
where the minimum mass for BH formation is simply assumed to be $30~\mathrm{M_\odot}$ 
(see, however, Heger et al.~\citeyear{Heger03} for a comprehensive discussion on the issue). 
From Yoon \& Langer~(\citeyear{Yoon06}). 
}\label{fig:fate}
\end{center}
\end{figure}

Not all stars which undergo quasi-homogeneous evolution end their life with a GRB. 
If the metallicity
is too high ($Z \ga Z_\odot/10$), mass loss results in a too strong spin-down of the star 
(Woosley \& Heger~\citeyear{Woosley05}; Yoon \& Langer~\citeyear{Yoon05}).  
At a given metallicity, homogeneously evolving stars of relatively 
low initial mass ($\sim 10\, M_{\odot}$) have a rather long CO core contraction time, 
and a rather massive He envelope after the He core exhaustion. In these stars, the CO core loses too much
angular momentum by magnetic torques after core helium burning 
to produce a GRB. This imposes a lower initial mass limit 
($M_\mathrm{min}$) for GRB formation. 
We find $M_\mathrm{min} \simeq 20~\mathrm{M_\odot}$ with slow semi-convection ($\alpha_\mathrm{SEM}=0.01$)
as discussed in Yoon \& Langer~(\citeyear{Yoon05}),
and $M_\mathrm{min} \simeq 12~\mathrm{M_\odot}$ with fast semi-convection ($\alpha_\mathrm{SEM}=1.0$),
in agreement with Woosley \& Heger~(\citeyear{Woosley05}). 

The upper initial mass limit for GRB formation is imposed by different
factors for different metallicities. 
At $Z=0.001$, stars with $M_\mathrm{init} \ga 40~\mathrm{M_\odot}$ experience
significant braking by rather strong mass loss, and cannot retain
enough angular momentum in the core. Therefore, only those stars
with $12~\mathrm{M_\odot} \la M_\mathrm{init} \la 40~\mathrm{M_\odot}$
are likely to produce GRBs, at $Z=0.001$. This upper mass limit will decrease
with increasing metallicity, due to stronger stellar wind mass loss. 
On the other hand, if the metallicity is very low ($Z \approx 10^{-5}$), 
the braking induced by mass loss is less important 
even for very massive stars.
However, homogeneously evolving stars with $M_\mathrm{init} \ga 60~\mathrm{M_\odot}$ 
yields CO cores  more massive than $40~\mathrm{M_\odot}$ and
likely undergo the pair-instability, 
preventing the formation of GRBs (Yoon \& Langer~\citeyear{Yoon05}). 

In Fig.~\ref{fig:fate}, we summarize the above discussion on the final fate of
massive stars at $Z=0.001$, in the mass-rotation plane, 
based on a grid of stellar models with different
initial masses and spin rates (about 40 model sequences; Yoon \& Langer~\citeyear{Yoon06}). 
Here, fast semi-convection ($\alpha_\mathrm{SEM}=1.0$) is adopted. 

\section{Properties of GRB progenitors}

\begin{figure}[!t]
\plottwo{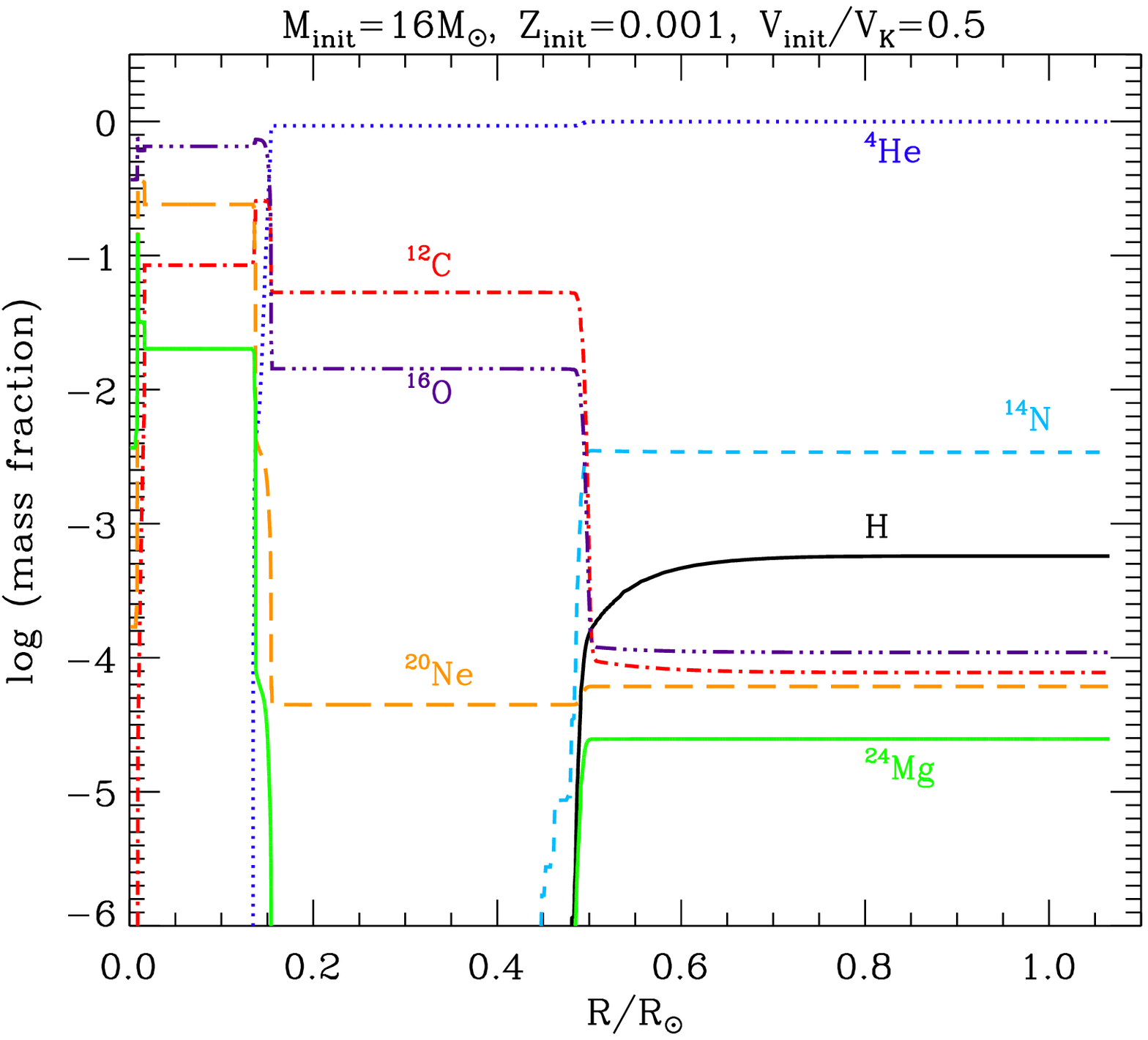}{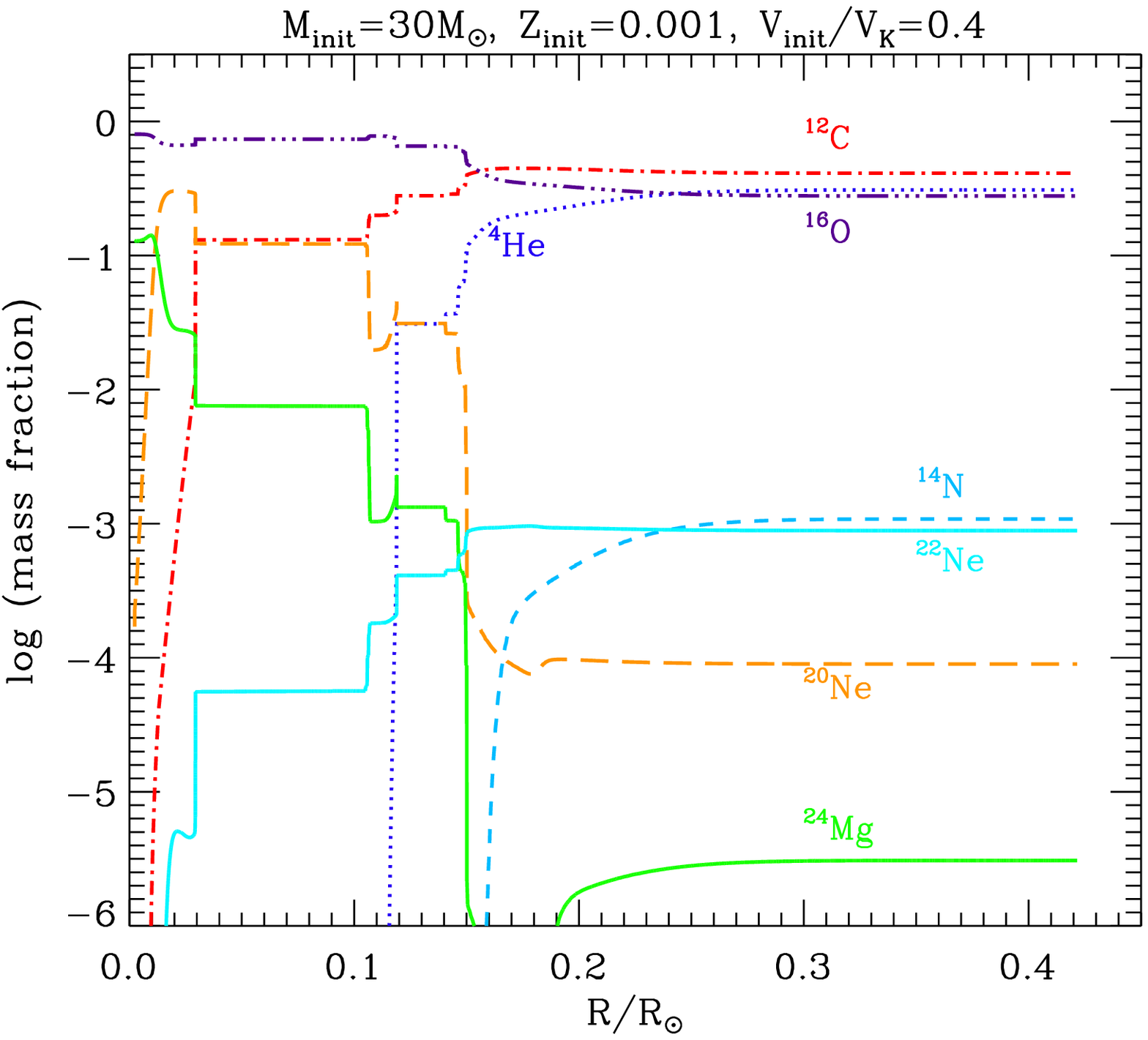}
\caption{Chemical structure of two GRB progenitor models, as examples for the
two different expected progenitor types. The left plot shows a "WN-type progenitor"
($ M_\mathrm{init} = 16~\mathrm{M_\odot}$, $V_\mathrm{init}/V_\mathrm{K}=0.5$),
with a final total mass of $13.07~\mathrm{M_\odot}$, a CO-core mass of $10.38~\mathrm{M_\odot}$, and hydrogen in its
outer $0.2~\mathrm{M_\odot}$. The right plot shows a "WC-type progenitor"
($ M_\mathrm{init} = 30~\mathrm{M_\odot}$, $V_\mathrm{init}/V_\mathrm{K}=0.4$),
which ends up with $20.08~\mathrm{M_\odot}$ and a CO-core mass of $16.88~\mathrm{M_\odot}$.
}\label{fig:chem}
\end{figure}

Our models at $Z=0.001$ predict two different types of GRB progenitors.
Initially less massive stars give less massive CO cores and more massive He envelopes.
As a result, stars with $12~\mathrm{M_\odot} \la M_\mathrm{init} \la 25~\mathrm{M_\odot}$
end their life as WN stars with rather massive helium-rich envelopes ($\Delta M_\mathrm{He} \ga 2.0~\mathrm{M_\odot}$). 
Those might be characterized as Type~Ib supernovae (Mazzali~2005, private communication; See Fig.~\ref{fig:fate}).
Remarkably, these stars also have hydrogen in their envelope (left panel in Fig.~\ref{fig:chem}), 
which might be relevant to the high velocity 
HI absorption line observed in the afterglow of GRB 021004 (Starling et al.~\citeyear{Starling05b}).
On the other hand, more massive stars become WC stars in the end, which will explode
as Type Ic supernovae (right panel of Fig.~\ref{fig:chem}). 
Our models also predict that GRB progenitors of "WC-type" are more compact, 
more massive, and have envelopes which are more enriched with $\alpha$-elements, than those of "WN-type" .

\section{Discussion}

The initial spin rate distribution of massive low-metallicity star is unknown, and thus 
our models can not readily predict a GRB formation rate at $Z=0.001$. 
However, Langer \& Norman (\citeyear{Langer06}; see also the General Discussion
after Session G, in this volume)
argue that if the majority of GRBs were restricted to metallicities below 
$Z = Z_\odot/10$, about 5 percent of all massive stars with such low
metallicity should produce a GRB in order to obtain GRBs at a rate
observed by BATSE. This requires that, at this and lower metallicity, more stars produce GRBs than
stars are predicted to die as WR star due to stellar wind mass loss
(cf. Meynet \& Maeder~\citeyear{Meynet05b}).
Consequently, a significant low-metallicity bias in GRBs would --- if confirmed --- 
not only be consistent with the quasi-chemically homogeneous evolution
scenario of GRB progenitors. It would require that indeed the evolution of 
low metallicity massive star differs significantly from that of massive
stars in our Galaxy, in support of the scenario outlined above.

\acknowledgements %%% Text of acknowledgements runs on after this command.
We are grateful to Alex de Koter, Paolo Mazzali, Philipp Podsiadlowski, and Ralph Wijers
for fruitful discussions. SCY is supported by the Netherlands Organization for Scientific
Research (NWO) through the VENI grant (639.041.406).

%%% THE BIBLIOGRAPHY
%%%
%%% CONSULT SECTION 3 OF "INSTRUCTIONS FOR AUTHORS" FOR HOW TO USE NATBIB.
%%% AUTHORS ARE ENCOURAGED TO USE EITHER THE "THEBIBLIOGRAPY" ENVIRONMENT
%%% BY UNCOMMENTING (DELETING THE "%" SYMBOL) THE COMMANDS BELOW, OR BY
%%% USING THE BIBTEX ENVIRONMENT. TO FIND OUT WHICH IS APPLICABLE TO YOUR
%%% CONTRIBUTION, CONSULT THE VOLUME EDITORS FOR YOUR PROCEEDINGS.
%%%

\end{document}